\documentclass[conference]{IEEEtran}
\IEEEoverridecommandlockouts
\usepackage{cite}
\usepackage{amsmath,amssymb,amsfonts}
\usepackage{algorithmic}
\usepackage{graphicx}
\usepackage{textcomp}
\usepackage{multirow}
\usepackage{makecell}
\usepackage{xcolor}
\def\BibTeX{{\rm B\kern-.05em{\sc i\kern-.025em b}\kern-.08em
    T\kern-.1667em\lower.7ex\hbox{E}\kern-.125emX}}
\begin{document}

\title{HTMD-Net: A Hybrid Masking-Denoising Approach to Time-Domain Monaural Singing Voice Separation
%\thanks{\textcolor{red}{obviously ePrev related research - hmmmm}}
}
%κατσε μην το ξεχασω αυτο κ παει πμ ετσι

\author{\IEEEauthorblockN{Christos Garoufis}
\IEEEauthorblockA{\textit{School of ECE, NTUA},
Athens, Greece \\
cgaroufis@mail.ntua.gr}
\and
\IEEEauthorblockN{Athanasia Zlatintsi}
\IEEEauthorblockA{\textit{School of ECE, NTUA},
Athens, Greece \\
nzlat@cs.ntua.gr}
\and
\IEEEauthorblockN{Petros Maragos}
\IEEEauthorblockA{\textit{School of ECE, NTUA},
Athens, Greece \\
maragos@cs.ntua.gr}}

\maketitle

\begin{abstract}
 The advent of deep learning has led to the prevalence of deep neural network architectures for monaural music source separation, with end-to-end approaches that operate directly on the waveform level increasingly receiving research attention. Among these approaches, transformation of the input mixture to a learned latent space, and multiplicative application of a soft mask to the latent mixture, achieves the best performance, but is prone to the introduction of artifacts to the source estimate. To alleviate this problem, in this paper we propose a hybrid time-domain approach, termed the HTMD-Net, combining a lightweight masking component and a denoising module, based on skip connections, in order to refine the source estimated by the masking procedure. Evaluation of our approach in the task of monaural singing voice separation in the musdb18 dataset indicates that our proposed method achieves competitive performance compared to methods based purely on masking when trained under the same conditions, especially regarding the behavior during silent segments, while achieving higher computational efficiency.
\end{abstract}

\section{Introduction}
\label{sec:intro}

Source separation is defined as the problem of decomposing an observed input signal into the components that constitute it. In the context of music processing, music source separation regards the isolation of vocal or instrumental tracks from a musical mixture. Historically, the problem of music source separation was tackled by signal processing-based methods \cite{virtanen2007,huang2012}. However, since the advent of deep learning, these methods have been gradually replaced by deep-learning based ones \cite{openunmix,stoller2018, pardo2018}. These methods can be split in two categories: Methods that operate in a time-frequency representation of the signal, usually its Short-Time Fourier Transform (STFT), in order to perform the separation procedure \cite{openunmix,jansson2017,takahashi2018}, and those that directly leverage the signal waveform to separate the desired sources in an end-to-end fashion \cite{stoller2018,grais2018,luo2019}.

The majority of deep learning approaches that operate in the STFT domain of a signal aim, inspired by traditional signal processing approaches, to predict a soft mask which, when applied via element-wise multiplication to the input magnitude spectrogram, will yield the magnitude spectrogram of the desired source \cite{jansson2017}. On the contrary, time-domain approaches can be split in two major categories: Autoencoder architectures with a number of skip connections that operate in multiple resolutions of the input waveform \cite{stoller2018,grais2018,defossez2019}, and architectures that follow the Encoder-Separator-Decoder paradigm \cite{luo2019}. In the second case, the encoder and the decoder are used to calculate an overcomplete latent mixture representation, upon which the separator calculates a mask to be applied, in a way akin to the STFT-based approaches.

Among time-domain approaches to audio source separation, neural network architectures based on the above-described Encoder-Separator-Decoder paradigm have achieved state-of-the-art performance in both speech separation \cite{luo2020} and music source separation \cite{defossez2019}. However, while music source separation approaches based on masking generally outperform those primarily utilizing skip connections, a drawback of these approaches regards the introduction of noise artifacts in the predicted source \cite{cano2018,drossos2018, defossez2019}. Previous works in the field \cite{drossos2018, park2018} that use an STFT-representation of the signal attempt to overcome this problem via refining the initial mask estimate by serially stacking either similar \cite{park2018} or suitably designed \cite{drossos2018} modules upon the initial masking network, and training the whole network in an end-to-end fashion. 

% tha borouse na bei sto fig me lexeis to mascing, denoising module? isos, hmm!! H iparxei xoros/xronos na ginei pio epexigimatiko k oxi toso abstract? isos :-/
\begin{figure*}[t]
    \begin{minipage}{\linewidth}
        \centering
        \centerline{\includegraphics[width=15cm]{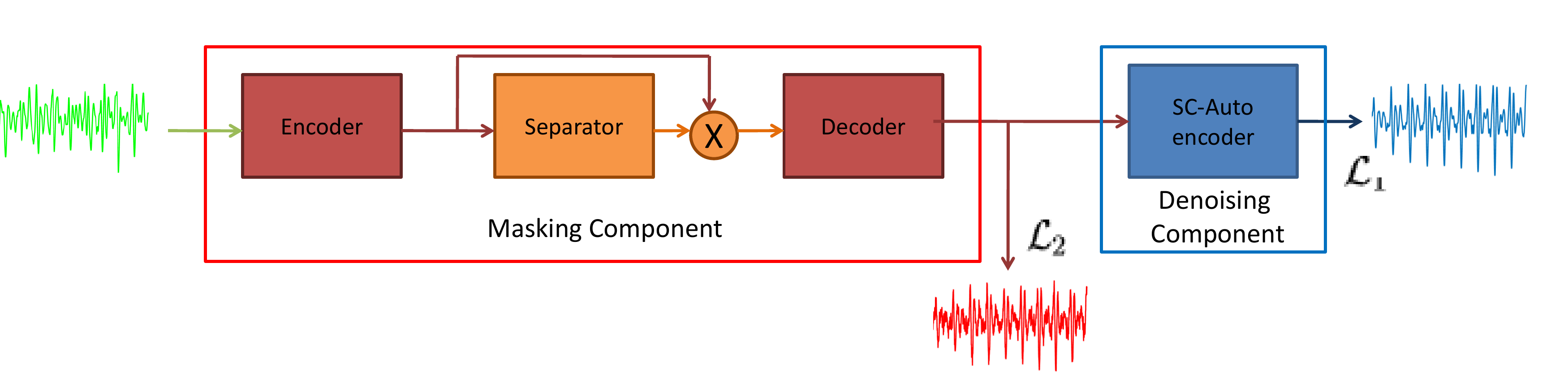}}
%        {outs_2.png}}
    \end{minipage}
    \vspace{-0.35cm}
    \caption{An overview of the whole system architecture, including the masking and denoising components and the separate losses over the intermediate and final source estimates.}
    \label{fig:ovrl}
    \vspace{-0.45cm}
\end{figure*}

In this work, we propose the HTMD-Net (short for Hybrid Temporal Masking-Denoising Network), a hybrid architecture for end-to-end monaural music source separation, consisting of two serially connected modules: one that provides an initial source estimate via applying a mask to an overcomplete learned latent representation of the mixture, and a second, based on multi-resolution analysis, in order to refine the initial estimate. While our approach shares the concept of serial module connection with \cite{drossos2018,park2018}, it deviates from those in that it operates directly in the time domain, as well as in the design of these modules. Furthermore, since the proposed framework allows the potential application of deep supervision, we conducted a number of experiments to gauge the behavior of HTMD-Net with respect to the training protocol used. %to design of these modules to anafero parapano... do we need repeating there?
% tora gia to design pou les.. elega pio poli na bei kati os motivation k oxi gia to design auto kat auto.. giati to ekanes design etsi?
%cg: reworded a bit the sentence above "... modules: one..."
Our proposed approach achieves competitive performance in the task of monaural singing voice separation in the widely used musdb18 \cite{rafii2017} dataset, compared to time-domain approaches that are purely based on either masking or multi-resolution analysis, especially regarding the alleviation of inter-source interferences, as well as higher computational efficiency. %The results are especially encouraging regarding the inter-source interference in the silent parts of the separated source. 

The rest of the paper is organized as follows: In Sec.~II, we outline the proposed architecture in detail. The experimental setup we utilize is described in Sec.~III, while in Sec.~IV we report and discuss our results. Finally, in Sec.~V we summarize our conclusions and propose some future research directions.

\section{Methodology}

%christos: keeping it 'a specific source' since it (hopefully) generalizes and is tested in singing voice separation.
The principal concept presented in this work regards decomposing the procedure of isolating
a specific source from a time-domain mixture into two separate subprocesses: 1) Finding an optimal mask to be applied on a latent mixture representation, as in \cite{luo2019}, and 2) refining the masked source estimate through a denoising module based on skip-connections. This two-step procedure is presented in Fig.~\ref{fig:ovrl}. %We shall now present the two basic building blocks of our system.

\subsection{Masking Network}

To perform the initial masking operation, we use a masking neural network, similar to \cite{luo2019}, that consists of a) a linear convolutional strided encoder that transforms the input mixture to a latent space, b) a mask estimation module, and c)  a linear decoder that consists of transposed convolutions, which reverts the estimated source to the time domain. In the original Conv-TasNet architecture \cite{luo2019}, the mask estimator is realised as a dilated Temporal Convolutional Network (dilated-TCN) and consists of $R$ modules, placed in succession. In turn, each module comprises of consecutive blocks that apply successively $1$x$1$ Convolutions and Depthwise Separable Convolutions with increasingly dilated kernels. Each convolutional block has two outputs, a mask estimate and a feature map to be used as input from the next block. The mask estimates from all blocks are summed together, and then scaled in a $[0,1]$ range via a sigmoid activation, in order to be multiplied with the encoder's output, thus yielding a representation of the desired source in the encoder's latent space. 

%i think i prefer it this way since the points of deviations from luo begin there.
In \cite{luo2019}, a total of 3 modules, each consisting of 8 convolutional blocks, were used, hence setting the maximum dilation rate to 8~\cite{luo2019}. In this work, we use only one module with a maximum dilation rate of 9, in order to increase its receptive field. Furthermore, after preliminary experiments, we replaced the PReLU activations and Layer Normalization operations with LeakyReLUs and Batch Normalization, respectively. 
The rest of the network's hyperparameters were left unchanged. 

\subsection{Denoising Network}

In order to refine the source estimate produced by the masking network, we serially attach a second trainable module to the masking component, so as to perform a denoising operation on it. We opted for an encoder-decoder network utilizing skip connections, since these architectures have proved efficient in the task of speech enhancement \cite{pascual2017, giri2019}. Thus, an architecture similar to the Wave-U-Net \cite{stoller2018} was used, replacing the convolutional bottleneck of the network with a recurrent module. The recurrent path consists of two bidirectional LSTM layers, as also proposed in \cite{defossez2019, kaspersen2020}, of 168 units each, and a LeakyReLU activation after the second layer. Also, in order to keep the computational costs low, the number of filters in both the encoder and the decoder were halved compared to the original implementation \cite{stoller2018}.  Otherwise, the structure of the network follows \cite{stoller2018}: %T
the encoder block consists of 1D-convolutional filters, followed by a LeakyReLU activation and a downsampling layer. Similarly, the decoder alternates between upsampling layers and 1D-convolutions, again followed by a LeakyReLU activation - with the exception of the output layer, which uses a tanh activation. The outputs of each encoder convolutional block pre-downsampling are concatenated with the feature maps of the respective decoder block after being upsampled via skip connections. %An overview of the architecture we use for the denoising module is presented in Fig.~2.

\subsection{Deep Supervision}

\begin{table*}[!h]

% song-wise metrics, segment-wise metrics
    \begin{center}
        \caption{Comparison of the HTMD-Net to a reimplementation of Conv-TasNet \cite{luo2019} and a Wave-U-Net \cite{stoller2018}. Bold denotes the best results at a statistical significance level of $p<0.01$. Higher values are better for all metrics, except PES (dB). }
        \vspace{-0.3cm}
    \begin{tabular}{|c|c||c|c|c||c|c|c|c|c|}\hline
    \multirow{2}{*}{Method} & \multirow{2}{*}{Loss  Function} & 
      \multicolumn{3}{c||}{Song-Wise Metrics} &
      \multicolumn{5}{c|}{Segment-Wise Metrics } \\ \cline{3-10}
    &  & SDR (dB) & SIR (dB) & SAR (dB) & SDR (dB) & SIR (dB) & SAR (dB) &PES (dB) & VAD (\%)  \\\hline
    HTMD-Net & (MSE, MSE) &5.16 & \textbf{10.24} & 8.53 & \textbf{4.69/0.60} & \textbf{9.80/6.98} & 7.92/6.53 & \textbf{-62.2} & \textbf{84.7} \\ \hline
     Conv-TasNet$^{*}$ & MSE &\textbf{5.25} & \textbf{9.74} & \textbf{8.85} & 4.83/-0.07 & 9.59/7.00 & \textbf{8.18/7.09} & -57.8 & 82.5   \\ \hline
     Wave-U-Net & MSE & 4.37 & 9.46 & 7.61 & 4.04/-0.14 & 9.00/6.49 & 7.17/6.24 & -61.4 & 82.1 \\ \hline \hline
         HTMD-Net & (MAE, MAE) &5.18 & \textbf{11.30} & 8.43 & 4.62/2.26 & \textbf{11.44/9.95} & 8.14/6.24 & \textbf{-80.1} & \textbf{85.3} \\ \hline
     Conv-TasNet$^{*}$ & MAE &\textbf{5.20} & 10.73 & \textbf {8.82} & \textbf{4.84/1.63} & 10.81/8.80 & \textbf{8.44/6.83} & -73.1 & 85.2   \\ \hline
     Wave-U-Net & MAE & 4.07 & 9.67 & 8.17 & 3.61/0.90 & 9.62/8.00 & 7.48/5.90 & -70.0 & 82.8 \\ \hline 
    \end{tabular}
    \label{tab:baselines}
    \end{center}
    \vspace{-0.5cm}
\end{table*}

Similar to \cite{drossos2018,park2018} we experiment with the application of deep supervision during training the network. Namely, we optimize the loss function:
%
%=\vspace{-0.3cm}
\begin{equation}
    \mathcal{L} = \alpha \mathcal{L}_1 + \beta \mathcal{L}_2,
    %\vspace{-0.3cm}
\end{equation}
where the losses $\mathcal{L}_1$, $\mathcal{L}_2$ correspond to the final and the intermediate source estimates, respectively, and $\alpha$, $\beta$ correspond to the loss weights. 

A potential advantage of the proposed deeply supervised framework regards the ability to incorporate different loss functions on the network's bottleneck and the final output. Since both $\mathcal{L}_1$, $\mathcal{L}_2$ are applied in the time domain, and not in a latent space, we experiment with using combinations of the mean square error (MSE) and mean absolute error (MAE) between the true and estimated sources as loss functions.

\section{Experimental Setup}

\begin{figure}[tb!]
\vspace{-0.5cm}
    \begin{minipage}{\linewidth}
        \centering
        \centerline{\includegraphics[width=7cm]{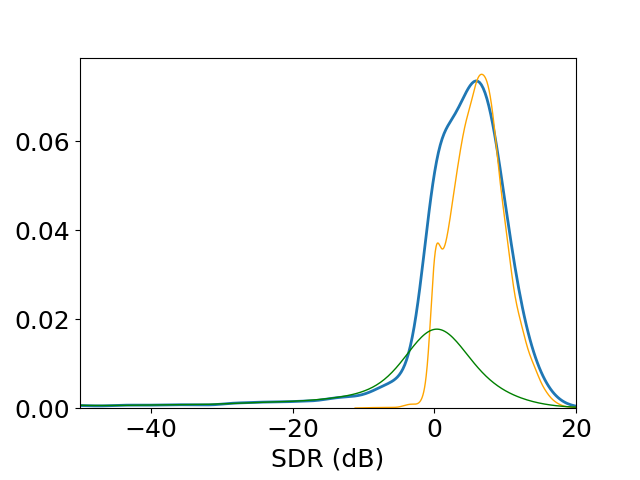}}
%        {outs_2.png}}
    \end{minipage}
    \caption{Kernel density estimate (KDE) for the segment-wise SDR for HTMD-Net, superimposed with the normalized KDEs of the segment-wise SDR corresponding to near-silent (green) or non-silent (orange) segments.}
    \label{fig:kde}
    \vspace{-0.4cm}
\end{figure}

\subsection{Dataset}

For our experiments in singing voice separation, we utilize the musdb18 \cite{rafii2017} dataset. This dataset consists of a total of 150 songs at stereo format and sampled at 44.1 kHz, as well as separate tracks for the vocals, bass, drums and the rest of the instrumental components (accompaniment) 
%why in parenthesis? an vgaleis tin parenthesi, vgazei noima? EPimeno se auto gia epomeno V
%hmmm as in rest of the accompaniment...
of each song, divided into a training set of 100 songs and a testing set of 50 songs. As preprocessing, we downsampled the audio excerpts corresponding to the song mixtures and the vocals to 22.05 kHz, after conversion from stereo to mono, as in~\cite{stoller2018}. 

\subsection{Training Setup and Baselines}

As baselines, we employ the following architectures:

\begin{itemize}
    \item A Conv-TasNet, consisting of 3 repetitions of the dilated-TCN separation module, as proposed originally in~\cite{luo2019}. The network's hyperparameters were set according to the optimal setup in~\cite{luo2019}, with the exception of Batch Normalization and LeakyReLU activations, while its input length was set to 16384 samples.  We note that we have used 9 dilated blocks in each repetition, in order to achieve a receptive field at least equal to the input length.
    \item A Wave-U-Net architecture \cite{stoller2018}, with the network's hyperparameters following the original implementation. %as well as a modified architecture (which we term Wave-U-Net+), in which we replace the CNN bottleneck with 2 bidirectional LSTM layers of 288 units each. 
\end{itemize}

All of the tested architectures were implemented in Keras and trained with the Adam optimizer \cite{kingma2014} with a learning rate of 0.0001, using a batch size of 16, with the exception of the Conv-TasNet, where we used a batch size of 8 due to memory limitations. All networks were trained using either the MSE or the MAE between the true and estimated sources as the loss function $\mathcal{L}_1$. 
The musdb18 training set was split in training and validation data, by using 75 out of the 100 songs for network training, and the rest as a validation set. No data augmentation was performed, and early stopping was applied after 20 epochs of no improvement in the validation set.

\subsection{Evaluation Protocol}

As our primary metric, we utilize the Signal-to-Distortion Ratio (SDR) between the true and predicted sources in the musdb18 test set, estimated over 1-sec segments. We further report, in accordance with \cite{vincent2006}, on the Signal-to-Artifact Ratio (SAR) and the Signal-to-Interference Ratio (SIR) - the former is used to gauge the existence of auditory artifacts, while the latter measures the contamination of the extracted sources. We report on both the song-wise median, as in \cite{sisec2018}, and the segment-wise median and mean, similar to \cite{stoller2018}.

\begin{table*}[h!]
    \begin{center}
    \caption{Comparison between the training protocols used for HTMD-Net. Higher values are better for all metrics, except PES (dB).
    %We report on the song-wise median SDR (dB), SIR (dB) and SAR (dB) values, the median and mean SDR (dB), SIR (dB) and SAR (dB) values over 1-sec segments, as well as the PES (dB) and VAD (\%) metrics.
    }
    \vspace{-0.2cm}
        % K kane mia tramba k sta dio tabs  to pio les proto sto cap k pio parousiazeis :-) to PES/VAD les einai se 4096-samle frames, antistoixei me to 1 secc segments? gi auto einai omadopiimena mazi? k edo ftiaxe sto tab k ti grammoula metaxi song k segment
    \begin{tabular}{|c|c||c|c|c||c|c|c|c|c|}\hline
   Loss Functions &  Loss Weights &
      \multicolumn{3}{c||}{Song-Wise Metrics} &
      \multicolumn{5}{c|}{Segment-Wise Metrics } \\ \cline{3-10}
      ($\mathcal{L}_2, \mathcal{L}_1)$ &$(\beta, \alpha)$ & SDR (dB)  & SIR (dB)  & SAR (dB)  & SDR (dB)  & SIR (dB) & SAR (dB)  &PES (dB) & VAD (\%)  \\\hline
     (MSE, MSE) & (0.5, 1) & 5.16 & 10.24 & 8.53 & 4.69/0.60 & 9.80/6.98 & 7.92/6.53 & -62.2 & 84.7 \\ \hline
    (MAE, MAE) & (0.5, 1) & 5.18 & \textbf{11.30} & 8.43 & 4.62/\textbf{2.26} & \textbf{11.44/9.95} & 8.14/6.24 & -80.1 & \textbf{85.3} \\ \hline
    (MAE, MSE) & (0.05, 1) & 5.16 & 10.33 & 8.36 & 4.68/0.34 & 9.97/7.87 & 8.06/6.65 & -59.9 & 84.2 \\ \hline
    (MSE, MAE) & (1, 0.1) & 5.21 & 11.29 & 8.34 & 4.74/2.21 & 10.90/9.03 & 7.95/6.04 & \textbf{-82.5} & 85.0 \\ \hline
    (-, MSE) & - & \textbf{5.30} & 10.05 & 8.62 & \textbf{4.76}/0.10 & 9.76/7.79 & \textbf{8.21/6.85} & -57.1 & 82.4 \\ \hline
    (-, MAE) & - & 4.77 & 9.88 & \textbf{8.63} & 4.37/1.88 & 9.58/8.02 & 7.94/6.43 & -74.5 & 84.8 \\ \hline
 
    \end{tabular}
    \label{tab:losses}
    \end{center}
    \vspace{-0.5cm}
\end{table*}

\begin{figure*}[tb!]
    \begin{minipage}{0.33\linewidth}
        \centering
        \centerline{\includegraphics[width=5.8cm]{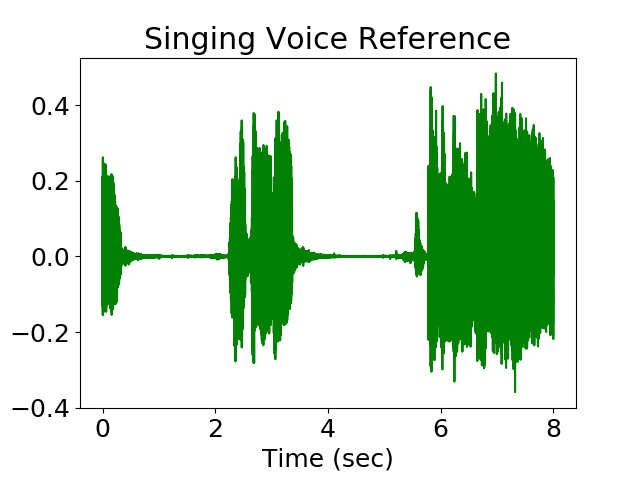}}
%        {outs_2.png}}
    \end{minipage}
        \begin{minipage}{0.33\linewidth}
        \centering
        \centerline{\includegraphics[width=5.8cm]{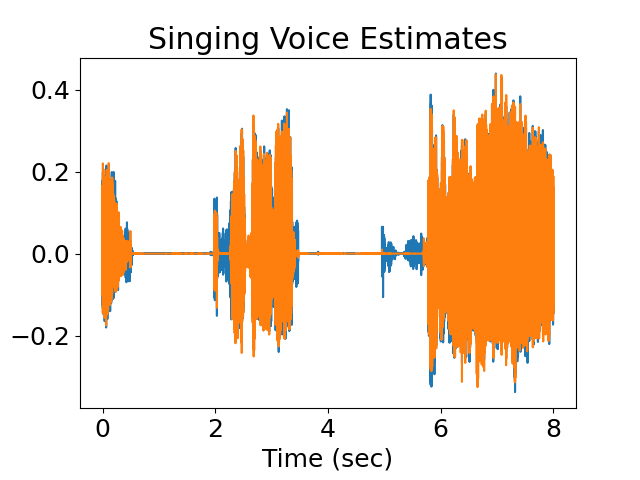}}
%        {outs_2.png}}
    \end{minipage}
    \begin{minipage}{0.33\linewidth}
        \centering
        \centerline{\includegraphics[width=5.8cm]{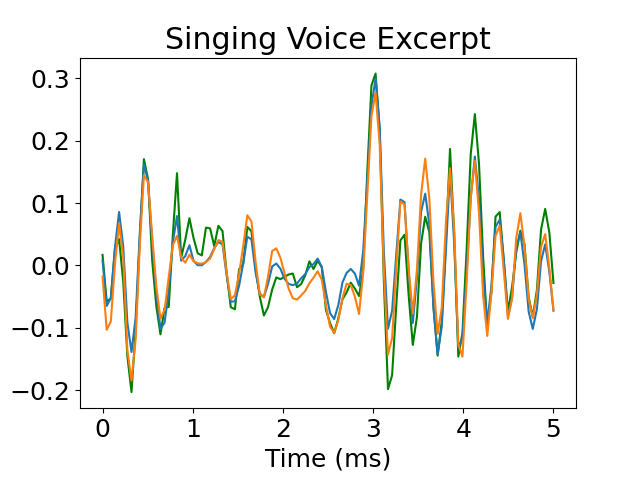}}
%        {outs_2.png}}
    \end{minipage}
    \vspace{-0.4cm}
    \caption{An 8-sec vocal track segment from the musdb18 test set, in green (left), the singing voice estimates for this segment provided by Conv-TasNet and HTMD-Net, in blue and orange respectively (center), and an utterance-level plot for both the reference and vocal estimates (right), using the same color code.}
    \label{fig:qualitative}
    \vspace{-0.4cm}
\end{figure*}

However, the above metrics are insufficient in assessing the performance of source separation algorithms in the time domain when used standalone, since they are not defined over silent segments of audio. Thus, we also employ as metrics the mean predicted energy at silence (PES), as in \cite{schulze2019, slizovskaia2020}, measured in 4096-sample frames, with a negative threshold of -100dB, and the correct vocal activity detection (VAD) percentage as measured in 20-ms frames of the network's output. To acquire the ground-truth VAD labels, we applied pyvad, a wrapper for the WebRTC Voice Activity Detection system, on the original vocal tracks. 

% sta parakato theoreis oti einai ok etsi? hmm, paizei na mas tin pei kapios pou den exei further details? theorontas pos ok, ta test einai gnosta k den exei refs px, ok, alla extra info?

\section{Results and Discussion}

%\begin{table*}[!h]
%    \begin{center}
%        \caption{Results of the ablation study conducted on HTMD-Net We report on the median and mean SDR (dB), SIR (dB) and SAR (dB) values over 1-sec segments, the song-wise median SDR (dB), SIR (dB) and SAR (dB) values, as well as the PES (dB) and VAD (\%) metrics. Bold denotes statistically significant results. }
%    \begin{tabular}{|c|c|c|c|c|c|}
    %\hline
%     Modification & SDR (median/mean, dB) & SI-SDR (median/mean, dB) & SAR (dB) &PES (dB) & VAD(\%) \\\hline
%     None & 4.69/0.61 & 3.12/-2.93  & 0 & -62.2 & 84.7 \\ \hline
%     (*)+ CNN Bottleneck & 4.56/0.63 & 3.04/-3.44 & 0 & -64.3 & 84.6 \\ \hline
    %(*) + Residual Connection & 4.64/-0.07  & 3.07/-3.24 &0 &-56.7 & 83.1 \\ \hline 
%     (*) + No Deep Supervision & 4.67/0.03 & 3.16/-4.06 &0& -58.6 & 82.9   \\ \hline 
%     (*) + Reverse Modules & 3.27/-2.23 & 1.24/-4.59 &0& -50.1 & 74.6 \\ \hline
%     Intermediate Output & 0/0 & 0/9 & 0 & 0 & 7 \\ \hline
 %   \end{tabular}
 %   \end{center}
 %   \vspace{-0.4cm}
%\end{table*}

%place table 5 there!

\textbf{Comparison to baselines:} The quantitative results of how HTMD-Net performs compared to the purely masking-based and skip-connection based baselines are presented in Table~\ref{tab:baselines}. We observe that our proposed architecture clearly outperforms the base Wave-U-Net \cite{stoller2018}, as well as performs comparably to our re-implementation of Conv-TasNet \cite{luo2019}. We note that the median SDR value corresponding to the Conv-TasNet is lower compared to that reported in~\cite{defossez2019}. This is likely due to a variety of factors, including the increased model size used in~\cite{defossez2019} or the stereo-mono conversion performed in our case. 

In order to assess whether the reported metric deviations between HTMD-Net and the two baselines could be attributed to random chance, pairwise statistical significance tests were performed for all metrics, among networks trained with the same training protocol. In specific, the paired Wilcoxon signed-rank test was performed over the distributions of all continuous metrics, and the paired McNemar's test over the binary variable denoting vocal activity estimation, using a p-value of $0.01$ in both cases. %We observe that HTMD-Net outperforms Wave-U-Net over all metrics. 
The results indicate that in comparison to the Conv-TasNet, HTMD-Net performs comparably considering the SDR, recording a lower median but a higher average value.
Additionally, it performs better in the absence of a vocal source, as it can be inferred from the lower PES and higher correct VAD percentage. We also note that in general, HTMD-Net records higher SIR scores, but lower SAR scores, in comparison to Conv-TasNet. This performance trend could be attributed to deep supervision, since multiple applications of the loss function should make the extracted source less contaminated by inter-source interferences. Finally, in comparison to the Wave-U-Net, the reported improvements of our approach are deemed statistically significant over all metrics.

%This also explains the higher mean SDR reported for HTMD-Net, 
By definition, the SDR is sensitive to outliers corresponding to near-silent segments~\cite{stoller2018}, which explains the big difference between the segment-wise median and mean SDR, and also the higher mean SDR reported for the HTMD-Net, since it performs better in near-silence. This effect is visualized in Fig.~\ref{fig:kde}, where the kernel density estimate (KDE) of the segment-wise SDR values is displayed for the HTMD-Net (blue), superimposed with the normalized KDEs of the subsets of the SDR values that correspond to near-silent (green) or non-silent (orange) 1-sec segments, as classified by pyvad. We observe that the vast majority of the outlier SDR values correspond to near-silent segments, since their SDR distribution almost overlaps with the overall one in negative SDR values. A similar trend was observed regarding SIR and SAR as well.

%The results indicate that our proposed method outperforms both baselines regarding SIR and the PES and VAD metrics, performs competitively to Conv-TasNet regarding SDR, while recording a lower SAR, on both segment and song level against it.

%% minor tora auto, to theto os skepsi..
% pano einai to comp to baselines, edo qualitative.. tha xorouse i lexi pano quantitative results / comparisons? K ISOS ta qualitative na pane kato apo ola ta quantitative??? 

% episis gia to fig isos na to eixame pei ksana? borei na ginei to portokali pio axno me kapio meiomeno opacity? :-/ na fainetai full apo kato to ble?

\textbf{Training Loss Schemes}: Upon inspection of Table~\ref{tab:baselines}, it is noted that using MAE as a loss function instead of the MSE does not necessarily imply improved network performance, but almost certainly provides more stable behavior regarding energy suppression in silent segments. Motivated by this, we also train variants of HTMD-Net using different loss functions in the bottleneck of the network and the final source estimate.

The results are reported in Table~\ref{tab:losses}, along with HTMD-Net variants trained without any deep supervision. We observe that among those variants, the model that was trained using the MSE as $\mathcal{L}_2$, and the MAE as $\mathcal{L}_1$, respectively, achieves competitive performance in most of the reported metrics.
% Ara auto to results den einai stat signif? tha borousame kapos na to deixoume omos? oxi me bold... kapos giati einai ontos to kalitero k to sizitas, alla me grigori matia ksefeugei apo to mati
% isos me italics?
We assume that with this loss function combination, the mask estimation module focuses more on following the vocal contour accurately, while the skip-connection module refines the initial estimation by enforcing silence in non-vocal segments.

% just that?? nothing more to say ? 
% ofc not, still wip 

 We further note that deep supervision has a significant effect in the quality of the network's output, especially regarding the Source-to-Interference Ratio (SIR), the mean segment-wise SDR values, and the silent segment performance as measured by PES. However, non-deeply supervised HTMD-Net variants achieve consistently good SAR values, and in the case of using the MSE loss, the song-wise median SDR is actually improved over our reimplementation of Conv-TasNet.

\begin{table}[h!]
    \vspace{-0.3cm}
    \begin{center}
      \caption{Comparison between the intermediate outputs in the bottleneck of the HTMD-Net, depending on the $\mathcal{L}_2$ used, when using the MAE as $\mathcal{L}_1$ .}
    \begin{tabular}{|c|c|c|c|}
    \hline
     $\mathcal{L}_2$ & SDR (dB) & SIR (dB) & SAR (dB) \\\hline
     MSE & \textbf{4.36}/-1.00 & \textbf{8.21/5.78} & 7.23/5.62     \\ \hline
     MAE & 4.09/\textbf{0.20} & \textbf{8.21}/5.29 & 7.79/7.06 \\ \hline
     - & -6.31/-13.1 & 2.81/1.22 & \textbf{7.26/7.25} \\ \hline
    \end{tabular}
    \label{tab:intermed}
    \end{center}
\vspace{-0.4cm}
\end{table}

\textbf{Behavior of Intermediate Output}: In Table~\ref{tab:intermed}, we report on the median and mean segment-wise SDR, SIR and SAR values for all HTMD-Net variants trained using the MAE as $\mathcal{L}_1$, measured at the bottleneck of the network where $\mathcal{L}_2$ was applied. We observe that while the deeply supervised variants record higher SDR and SIR values at the bottleneck, in the case where no $\mathcal{L}_2$ was applied, the reported SAR values are competitive, despite the lack of any supervision at this point. Given the overall performance of the non-deeply supervised variants, these results merit further exploration.

\textbf{Qualitiative Results\footnote{Audio samples/code available at: https://github.com/cgaroufis/HTMD-Net}
}: In Fig.~\ref{fig:qualitative} 
(left), we present an 8-second segment of the track ``Secretariat - Over the Top'' from the musdb18 test set. We can see that
this segment contains two silent sections at 1 and 4 sec. From Fig.~\ref{fig:qualitative}  (center), we can deduce that the performance of Conv-TasNet (blue) significantly deteriorates in the silent sections, whereas HTMD-Net (orange) is more successful in removing the other active instrumental sources in these areas. On the other hand, the vocal estimate provided by Conv-TasNet is closer to the reference vocals (green) regarding the utterance-level vocal peaks, as inferred from Fig.~\ref{fig:qualitative} ~(right). These observations agree with the quantitiative
results presented earlier, since HTMD-Net achieves slightly lower median SDR compared to the Conv-TasNet, but better performance at silent sections as measured by VAD (\%), PES, as well as the mean SDR.

% kapou na bei k to git gia ta akoustika simata pou exeis anevasei! me footnote mallon
%roger

\textbf{Runtime Comparison:} Finally, in Table~\ref{tab:footprint}, the total model sizes for Conv-TasNet, Wave-U-Net, and HTMD-Net are presented, along with the required time (in sec) to process 30 sec of audio, sampled at 22.05 kHz, in an AMD-A9 CPU and an NviDIA Ge-Force GTX 1080 GPU, respectively, averaged over 5 runs. We note that the HTMD-Net has a marginally smaller model size compared to our Conv-TasNet adaptation, and below half the size of a Wave-U-Net. The processing time is higher than the one recorded for the Wave-U-Net, but significantly less compared the one of Conv-TasNet, and approaches real-time performance even on the AMD A9 CPU.

\begin{table}[h!]
    \vspace{-0.3cm}
    \begin{center}
      \caption{Comparison of the HTMD-Net to a reimplementation of Conv-TasNet \cite{luo2019} as well as a Wave-U-Net \cite{stoller2018}, regarding execution runtime and parameter footprint.}
    \begin{tabular}{|c|c|c|c|}
    \hline
     Method & CPU-time & GPU-time & \# Params \\\hline
     Conv-TasNet$^{*}$ & 140.7& 0.65 & 5.5M     \\ \hline
     Wave-U-Net & 13.6 & 0.07 & 10.3M\\ \hline
     HTMD-Net & 50.5 & 0.14 & 4.5M  \\ \hline
    \end{tabular}
    \label{tab:footprint}
    \end{center}
\vspace{-0.4cm}
\end{table}

\section{Conclusions}

In this paper, we presented a hybrid approach to monaural singing voice separation that employs both a masking component in order to find an optimal separating mask and a denoising module with skip connections to further reduce the inter-source interference artifacts introduced by the masking procedure. The results of our method are promising, since HTMD-Net is able to perform competitively with the best-performing time-domain architectures when trained under similar settings, achieving a more stable behavior in silent sections, while maintaining a smaller parameter footprint and requiring less time for inference. In the future, we are interested in whether our findings can scale to larger input lengths, or time-domain adaptations of architectures that are designed to handle multiple separable sources \cite{meseguer2019, kadandale2020}. Furthermore, perceptual subjective evaluation tests could be performed, in order to support the objective results presented in this work. %while further experimentation with deep supervision in order to apply additional constraints in the latent masks or the embeddings of the denoiser is a promising research direction as well.

\bibliographystyle{IEEEtran}
\bibliography{refs}

\end{document}